\documentclass[10pt,conference]{IEEEtran}

\usepackage{algorithm,
            algorithmic,
            alltt,
            amsmath,
            array,
            balance,
            booktabs,
	        cite,
	        expl3,
	        float,
	        flushend,
	        framed,
	        graphicx,
	        lipsum,
	        listings,
	        makecell,
	        mathtools,
	        multirow,
	        pgfplots,
	        pgfplotstable,
	        pifont,
	        scalefnt,
	        soul,
	        subfig,
	        syntax,
	        tabularx,
	        tcolorbox,
	        threeparttable,
	        tikz,
	        underscore,
	        url,
	        verbatim,
	        xcolor
}

\usepackage[export]{adjustbox}
\usepackage{paralist}
\usetikzlibrary{patterns}

\ExplSyntaxOn
\newcommand\latinabbrev[1]{
  \peek_meaning:NTF . {
    #1\@}%
  { \peek_catcode:NTF a {
      #1., \@ }%
    {#1., \@}}}
\ExplSyntaxOff

\newcommand{\CASE}[1]{\STATE \textbf{case} #1\textbf{:} \begin{ALC@g}}
\newcommand{\ENDCASE}{\end{ALC@g}}

\newcommand{\DEFAULT}{\STATE \textbf{default:} \begin{ALC@g}}
\newcommand{\ENDDEFAULT}{\end{ALC@g}}
\newcommand{\DEFAULTLINE}[1]{\STATE \textbf{default:} }

\newsavebox{\supbox}
\newcommand{\bsup}{\begin{lrbox}{\supbox}$\tt\scriptstyle}
\newcommand{\esup}{$\end{lrbox}{}^{\usebox{\supbox}}}

\usetikzlibrary{shapes.gates.logic.US,trees,positioning,arrows}
\usetikzlibrary{shapes.multipart}
\usetikzlibrary{shapes,arrows}
\usetikzlibrary{arrows,shapes,backgrounds}

\tikzstyle{vertex}=[ellipse,fill=black!25,minimum size=20pt, inner sep=0pt]
\tikzstyle{edge} = [draw,thin,-]
\tikzstyle{glabel} = [text width=1.5cm,text centered,font=\bf]

\pgfdeclarelayer{bg}    
\pgfsetlayers{bg,main}

\pgfplotsset{compat=1.9}

\let\footnotesize\scriptsize

\definecolor{lightpurple}{rgb}{0.8,0.8,1}
\definecolor{codebg}{RGB}{255,255,255}
\definecolor{commentcolor}{RGB}{11,140,11}
\pgfplotsset{
    discard if/.style 2 args={
        x filter/.code={
            \edef\tempa{\thisrow{#1}}
            \edef\tempb{#2}
            \ifx\tempa\tempb
                
            \fi
        }
    },
    discard if not/.style 2 args={
        x filter/.code={
            \edef\tempa{\thisrow{#1}}
            \edef\tempb{#2}
            \ifx\tempa\tempb
            \else
                
            \fi
        }
    }
}

\usepackage[normalem]{ulem} 
\usepackage{xcolor}
\usepackage{ifthen}
\usepackage{enumitem}
\newboolean{showcomments}

\usepackage[numbers]{natbib}
\usepackage{enumitem}
\usepackage{amssymb}

\setboolean{showcomments}{true}

\ifthenelse{\boolean{showcomments}}
{\newcommand{\nbc}[3]{
 {\colorbox{#3}{\bfseries\sffamily\scriptsize\textcolor{white}{#1}}}
 {\textcolor{#3}{\sf\small$\blacktriangleright$\textit{#2}$\blacktriangleleft$}}
 }
}
{\newcommand{\nbc}[3]{}
 }

\definecolor{black}{HTML}{000000}
\definecolor{red}{HTML}{FF0000}

\definecolor{light-gray}{gray}{0.9}
\sethlcolor{light-gray}
\definecolor{bole}{rgb}{0.47, 0.27, 0.23}
\definecolor{byzantium}{rgb}{0.44, 0.16, 0.39}

\begin{document}


\title{Investigating Technology Usage Span by Analyzing Users' Q\&A Traces in Stack Overflow}

\author{\IEEEauthorblockN {Saikat Mondal, Debajyoti Mondal, Chanchal K. Roy }
\IEEEauthorblockA{Department of Computer Science, University of Saskatchewan, Canada\\ }
{\{saikat.mondal, d.mondal, chanchal.roy\}}@usask.ca
}





\maketitle

\begin {abstract} 
Choosing an appropriate software development technology (e.g., programming language) is challenging due to the proliferation of diverse options. The selection of inappropriate technologies for development may have a far-reaching effect on software developers' career growth. Switching to a different technology after working with one may lead to a complex learning curve and, thus, be more challenging. Therefore, it is crucial for software developers to find technologies that have a high usage span. Intuitively, the usage span of a technology can be determined by the time span developers have used that technology.
Existing literature focuses on the technology landscape to explore the complex and implicit dependencies among technologies but lacks formal studies to draw insights about their usage span. This paper investigates the technology usage span by analyzing the question and answering (Q\&A) traces of Stack Overflow (SO), the largest technical Q\&A website available to date. In particular, we analyze 6.7 million Q\&A traces posted by about 97K active SO users and see what technologies have appeared in their questions or answers over 15 years. According to our analysis, C\# and Java programming languages have a high usage span, followed by JavaScript. 
Besides, developers used the .NET framework, iOS \& Windows Operating Systems (OS), and SQL query language for a long time (on average). 
Our study also exposes the emerging (i.e., newly growing) technologies. For example, usages of technologies such as SwiftUI, .NET-6.0, Visual Studio 2022, and Blazor WebAssembly framework are increasing. The findings from our study can assist novice developers, startup software industries, and software users in determining appropriate technologies. This also establishes an initial benchmark for future investigation on the use span of software technologies.
\end {abstract}




\IEEEpeerreviewmaketitle

\section{Introduction}
\label{sec:introduction}

Software development technologies have advanced dramatically over the past few decades. A diverse set of technologies are available 
today to perform similar tasks, and they are evolving rapidly over time \cite{chen2016mining}. Here, the term \emph{technology} broadly refers to programming languages, frameworks, platforms, libraries, OS, and Database Management Systems (DBMS) in the context of software engineering. However, choosing the appropriate technology (e.g., programming language) is incredibly challenging amid hundreds of popular options. Inappropriate technology selection may -- (1) create career challenges, (2) waste developers' precious time, and (3) cause an unwanted switch that adds an extra burden to learning new technology with a complex learning curve. Developers thus often seek technologies with a high usage span on which they can build their careers and continue working for a long time. However, automatically discovering technologies with high \emph{usage span} requires reliable data from software developers' prolonged activity. In this study, usage span refers to how long developers prefer to work with a technology.

\begin{figure}[pt]
	\centering
	\includegraphics[width = 2.5in]{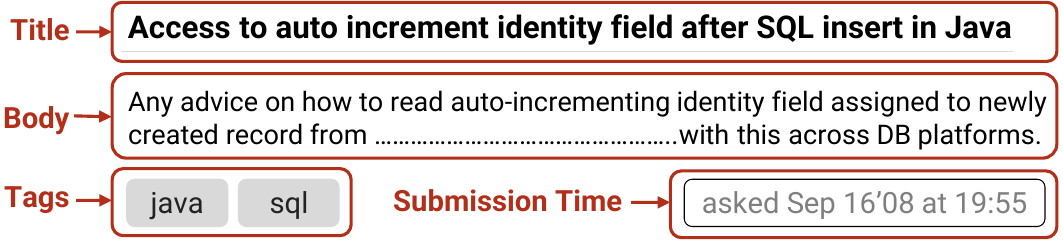}
	\caption{A 
	question 
	in SO with specified technology \cite{technologyExample}.}
	\label{fig:technology-example}
	\vspace{-4mm}
\end{figure}

Stack Overflow (SO) is the largest and most popular technical Q\&A website. It accumulates the knowledge and expertise of about 21 million developers from all over the world \cite{api}. This knowledge reflects insight into the development activities of those developers \cite{zou2015non}. 
Several existing studies exploit this crowd-sourced knowledge from SO to investigate the technology landscape \cite{chen2016mining, chen2016techland}, trends \cite{woon2009framework, barua2014developers, morris2002diva, kim2008visualization} and developers' skills \cite{montandon2020skills}.
For example, Chen et al.~\cite{chen2016mining, chen2016techland} attempt to mine the technology landscape utilizing SO question tags. They focus on exposing the complex relationships among technologies to offer an aggregated view of associated technologies but do not study the span or span of those aggregated technologies. 
Montandon et al.~\cite{montandon2020skills} analyze 20K job opportunities 
to characterize the hard and soft skills required in IT companies. Barua et al.~\cite{barua2014developers} attempt to mine the popular topics (e.g., jQuery) discussed in SO. Note that popularity does not guarantee usage span.  
We thus notice a lack of research that carefully investigates the usage span of technologies.

This study investigates about 97K developers' activity footprints in SO over 15 years. In particular, we analyzed more than 6.7 million questions and answers submitted by those developers to estimate the usage span of technologies. We hypothesized that (1) developers post questions or submit answers on those technologies they actively work with, and (2) if the average working year on a particular technology is high, the usage span of that technology is high, and vice versa. In SO, the technologies (e.g., Java) discussed in the questions are mentioned as tags (e.g., Fig. \ref{fig:technology-example}). We  
utilize the tags from questions to find the technologies and estimate their usage span. In particular, we answer two research questions and thus make two contributions to this study.


    \noindent \textbf{RQ$_1$) Can we leverage the developers' Q\&A footprints in Stack Overflow to reveal and compare the usage span of various software development technologies?} 
    We first measure how many years each of the selected developers of SO used a particular technology (i.e., usage year). 
    Second, we count the total usage years of technology and the number of developers who use that technology. 
    Then, we measure the average usage years of each technology, dividing the total usage years by the total number of developers.
    We report the top 50 technologies with a high usage span. It comprises 17 programming languages (e.g., JavaScript, C\#), 10 software frameworks (e.g., asp.net), five operating systems (e.g., iOS), one development platform (e.g., Azure), one version control system (e.g., git),  two IDE (e.g., visual-studio), five database \& database management systems (e.g., SQL Server), two query languages (e.g., SQL), two compilers (e.g., GCC) and five others (e.g., Google Services).

    \noindent \textbf{RQ$_2$) Can we unveil emerging technologies to which more developers are switching over time?} We determine the emerging (i.e., newly growing) technologies by analyzing whether the number of users is increasing over the years. Then, we report 20 technologies that more users have
    used with time, such as SwiftUI, .NET-6.0 \& Blazor WebAssembly frameworks, and Visual Studio 2022 IDE.
    \smallskip

\noindent Our study findings could help to select/recommend technologies with high usage span for (1) novice developers, (2) startup software companies, (3) software clients, (4) educators in computer science (or information technology), and (5) software engineering researchers.

\smallskip
\noindent\textbf{Replication Package.} Replication package that contains our data \& source code can be found in our online appendix \citep{replicationPackage}.

\begin{figure}[!htb]
	\centering
	\includegraphics[width = 3.45in]{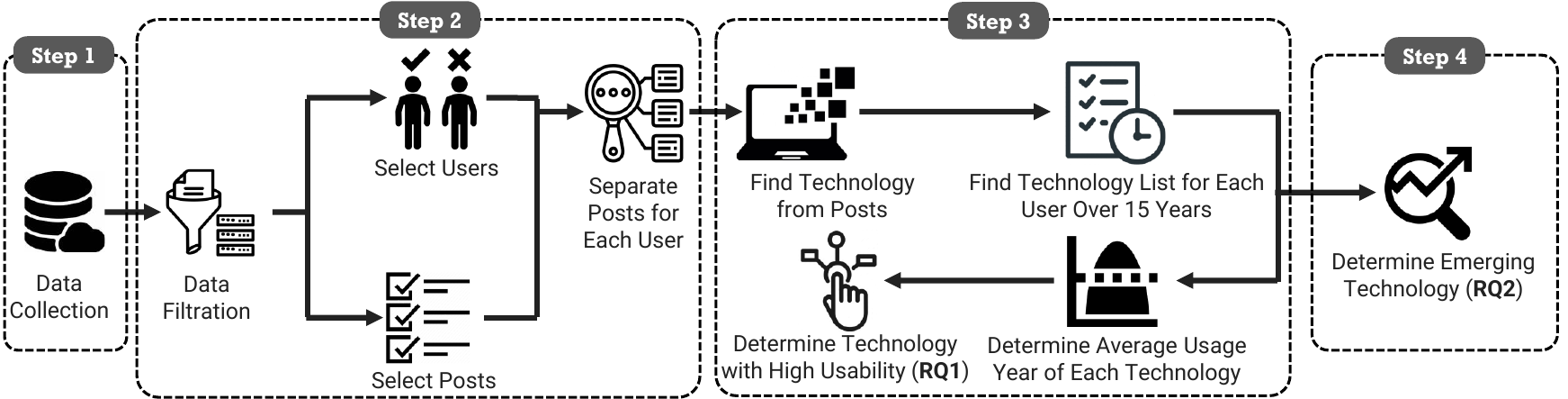}
	\caption{Schematic diagram of our exploratory study.}
	\label{fig:schematic-diagram}
	\vspace{-3mm}
\end{figure}

\begin{table*}[!htb]
\centering
\caption{Users technology usage over the years}
\label{table:technology-usage-per-user}
\resizebox{6.5in}{!}{%
\begin{tabular}{l|p{2.5cm}|p{3cm}|p{1.5cm}|p{1.9cm}|p{1.5cm}|p{0.8cm}|p{1cm}|p{2cm}|p{2cm}|p{2cm}|p{1.5cm}|p{0.7cm}|p{0.7cm}|p{1cm}|p{1.5cm}} \toprule
\multirow{2}{*}{\textbf{User ID}} & \multicolumn{15}{c}{\textbf{Technology}} \\ \cmidrule{2-16}
                                  & \textbf{2008} & \textbf{2009} & \textbf{2010} & \textbf{2011} & \textbf{2012} & \textbf{2013} & \textbf{2014} & \textbf{2015} & \textbf{2016} & \textbf{2017} & \textbf{2018} & \textbf{2019} & \textbf{2020} & \textbf{2021} & \textbf{2022}\\ \midrule
\multirow{3}{*}{\textbf{56}}   
& c\# css asp.net .net visual-studio html windows ajax javascript sql 
& c\# jaxb css java safari asp.net .net visual-studio html netbeans javascript macos
& asp.net .net visual-studio javascript
& c\# oracle asp.net .net windows sql-server javascript
& asp.net sql-server
& c\# css html
& html ios javascript
& c\# oracle asp.net visual-studio entity-framework
& python postgresql applescript macos
& python postgresql applescript macos
& github git macos javascript
& python
& macos
& postgresql
& google-service \\ \bottomrule 

\end{tabular}
}
\end{table*}

\section{Methodology}

A schematic diagram of our conducted exploratory study is shown in Fig. \ref{fig:schematic-diagram}.  We first downloaded the August 2023 SO data dump. We then select our target users and their posts over $15$ years. Next, we analyze the technologies associated with the posts to find the ones with a high usage span or growing usages over time (i.e., emerging).

\textbf{Data Collection.} We collected the August 2023 data dump of SO from the Stack Exchange website \cite{datadumpapi}. This contains all the essential information about questions, answers, users, and tags (i.e., technology) of the SO Q\&A website.

\textbf{Data Preprocessing.} We selected the users who created their accounts in SO in 2008 and 2009. We found a total of $96,786$ ($20,966$ in 2008 + $75,820$ in 2009) active users. We selected these users to get Q\&A traces over a long period (e.g., 15 years). For example, we can get Q\&A traces for a maximum of seven years (till 2023) if the user created an account in 2015. We then find the selected users' Q\&A footprints in SO for over 15 years. In particular, we extract Q\&A footprints -- (1) from 2008 to 2022 (i.e., 15 years) for the users who created their accounts in 2008 and (2) from 2009 to 2023 (i.e., 15 years) for those who created their accounts in 2009. Our dataset contains $6,715,748$ posts ($1,388,658$ questions + $5,327,090$ answers) that were submitted by our selected users. We separate the posts for each of the users.

\begin{figure*}[!htb]
\centering
      \resizebox{6in}{!}{%
      \begin{tikzpicture}
        	\begin{axis}[
        	xtick={1,2,3,4,5,6,7,8,9,10,11,12,13,14,15,16,17,18,19,20,21,22,23,24,25,26,27,28,29,30,31,32,33,34,35,36,37,38,39,40,41,42,43,44,45,46,47,48,49,50},
        	xticklabels={
                        javascript,
                        c\#,
                        java,
                        python,
                        c++,
                        php,
                        html,
                        css,
                        c,
                        ruby,
                        delphi,
                        objective-c,
                        r,
                        swift,
                        haskell,
                        scala,
                        f\#,
                        asp.net,
                        .net,
                        ruby-on-rails,
                        spring,
                        entity-framework,
                        django,
                        cocoa,
                        qt,
                        angular,
                        laravel,
                        ios,
                        windows,
                        android,
                        linux,
                        macos,
                        azure,
                        git,
                        visual-studio,
                        eclipse,
                        sql-server,
                        mysql,
                        database,
                        oracle,
                        postgresql,
                        sql,
                        tsql,
                        gcc,
                        visual-c++,
                        google-service,
                        iphone,
                        apache,
                        amazon-service,
                        iis
                        },
        	x tick label style={rotate=90,anchor=east},
        	enlarge y limits=false,
        	enlarge x limits=0.02,
        	ymin=0,ymax=5.5,
        	ybar,
        	width=8in,
        	height = 2in,
        	ytick={1,2,3,4,5},
            yticklabels={1,2,3,4,5},
        	ylabel={Avg. usage span (in years)},
        	ymajorgrids=false,
        	major x tick style = {opacity=0},
        	minor x tick num = 1,    
        	minor tick length=1ex,
        	legend style={
            	font=\small,
            	legend cell align=left,
                anchor=north,
                legend columns=1,
                at={(1.22,0.99)}
            },
            nodes near coords style={rotate=90,  anchor=west, font=\small},
        	nodes near coords =\pgfmathprintnumber{\pgfplotspointmeta},
        	every axis plot/.append style={
            ybar,
            bar width=.2cm,
            bar shift=0pt,
            fill}
        	]
            \addlegendimage{red!60}
            \addlegendimage{green!60}
            \addlegendimage{purple!60}
            \addlegendimage{blue!60}
            \addlegendimage{pink!60}
            \addlegendimage{black!60}
            \addlegendimage{brown!60}
            \addlegendimage{byzantium!60}
            \addlegendimage{yellow!80}
            \addlegendimage{orange!60}
            
            \addplot[draw=red!60, fill=red!60]coordinates {(1,4.4)};
            \addplot[draw=red!60, fill=red!60]coordinates {(2,3.9)};
            \addplot[draw=red!60, fill=red!60]coordinates {(3,3.2)};
            \addplot[draw=red!60, fill=red!60]coordinates {(4,3.1)};
            \addplot[draw=red!60, fill=red!60]coordinates {(5,2.5)};
            \addplot[draw=red!60, fill=red!60]coordinates {(6,2.5)};
            \addplot[draw=red!60, fill=red!60]coordinates {(7,2.5)};
            \addplot[draw=red!60, fill=red!60]coordinates {(8,2.1)};
            \addplot[draw=red!60, fill=red!60]coordinates {(9,2.1)};
            \addplot[draw=red!60, fill=red!60]coordinates {(10,2)};
            \addplot[draw=red!60, fill=red!60]coordinates {(11,2)};
            \addplot[draw=red!60, fill=red!60]coordinates {(12,1.9)};
            \addplot[draw=red!60, fill=red!60]coordinates {(13,1.8)};
            \addplot[draw=red!60, fill=red!60]coordinates {(14,1.8)};
            \addplot[draw=red!60, fill=red!60]coordinates {(15,1.7)};
            \addplot[draw=red!60, fill=red!60]coordinates {(16,1.7)};
            \addplot[draw=red!60, fill=red!60]coordinates {(17,1.6)};

            \addplot[draw=green!60, fill=green!60]coordinates {(18,2.9)};
            \addplot[draw=green!60, fill=green!60]coordinates {(19,2.8)};
            \addplot[draw=green!60, fill=green!60]coordinates {(20,2.1)};
            \addplot[draw=green!60, fill=green!60]coordinates {(21,1.9)};
            \addplot[draw=green!60, fill=green!60]coordinates {(22,1.9)};
            \addplot[draw=green!60, fill=green!60]coordinates {(23,1.9)};
            \addplot[draw=green!60, fill=green!60]coordinates {(24,1.7)};
            \addplot[draw=green!60, fill=green!60]coordinates {(25,1.6)};
            \addplot[draw=green!60, fill=green!60]coordinates {(26,1.6)};
            \addplot[draw=green!60, fill=green!60]coordinates {(27,1.6)};
            \addplot[draw=purple!60, fill=purple!60]coordinates {(28,2.2)};
            \addplot[draw=purple!60, fill=purple!60]coordinates {(29,2.2)};
            \addplot[draw=purple!60, fill=purple!60]coordinates {(30,2.1)};
            \addplot[draw=purple!60, fill=purple!60]coordinates {(31,1.9)};
            \addplot[draw=purple!60, fill=purple!60]coordinates {(32,1.7)};
            \addplot[draw=blue!60, fill=blue!60]coordinates {(33,1.8)};
            \addplot[draw=pink!60, fill=pink!60]coordinates{(34,1.8)};
            \addplot[draw=black!60, fill=black!60]coordinates{(35,2.4)};
            \addplot[draw=black!60, fill=black!60]coordinates{(36,1.6)};
            \addplot[draw=brown!60, fill=brown!60]coordinates {(37,2.2)};
            \addplot[draw=brown!60, fill=brown!60]coordinates {(38,2.1)};
            \addplot[draw=brown!60, fill=brown!60]coordinates {(39,1.9)};
            \addplot[draw=brown!60, fill=brown!60]coordinates {(40,1.7)};
            \addplot[draw=brown!60, fill=brown!60]coordinates {(41,1.7)};
            \addplot[draw=byzantium!60, fill=byzantium!60]coordinates {(42,2.3)};
            \addplot[draw=byzantium!60, fill=byzantium!60]coordinates {(43,1.7)};
            \addplot[draw=yellow!80, fill=yellow!80]coordinates {(44,1.6)};
            \addplot[draw=yellow!80, fill=yellow!80]coordinates {(45,1.6)};
            \addplot[draw=orange!60, fill=orange!60]coordinates {(46,2.0)};
            \addplot[draw=orange!60, fill=orange!60]coordinates {(47,1.8)};
            \addplot[draw=orange!60, fill=orange!60]coordinates {(48,1.7)};
            \addplot[draw=orange!60, fill=orange!60]coordinates {(49,1.6)};
            \addplot[draw=orange!60, fill=orange!60]coordinates {(50,1.6)};
            
            \legend	{Programming/Scripting/Style Sheet/Markup Language,
            		 Software Framework,
            		 Operating System,
            		 Development Platform,
            		 Version Control System,
            		 Integrated Development Environment,
            		 Database/Database Management System,
            		 Query Language,
                      Compiler,
            		 Other
            		 }
        	\end{axis}
    	\end{tikzpicture}
    	}
    	
\caption{Average usage span of technologies.}
\vspace{-3mm}
\label{fig:technology-usability}
\end{figure*}
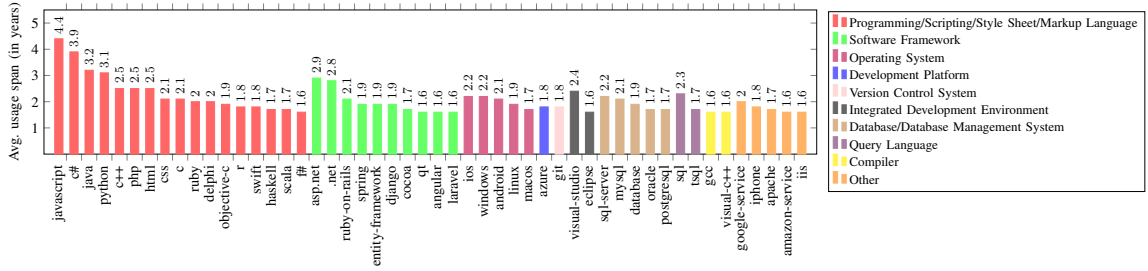

\textbf{Technology Selection.}
We attempt to capture the standard technologies and calculate their usage span (\textbf{RQ\textsubscript1}). In SO, tags often capture the technology \cite{mondal2021early}. SO data dump contains $65,152$ tags. It stores how many questions (i.e., count) used each tag. First, we sorted the tags according to their count in descending order. We then select the top 1K tags that are most often used in questions.  
Second, we manually analyze the description of these tags. We aim to find tags that represent standard software technologies, such as programming language, software framework \& platform, libraries, Database Management Systems (DBMS),  Integrated Development Environment (IDE), and Operating Systems (OS). Our manual analysis selects 185 tags. Among 185, 25 were libraries. We then merge these libraries with corresponding programming languages. For example, jQuery is a library of JavaScript. We thus merge jQuery with JavaScript. Finally, we get 160 tags that represent standard technology (online appendix \cite{replicationPackage} contains full list).
However, we consider all the tags (e.g., $65,152$) to capture emerging technologies (\textbf{RQ\textsubscript2}). Here, we attempt to see which version or component of standard technology has gained more attention from the developer community.

\textbf{Determining High Usage Technologies.} 
We first separated the list of posts for each of our selected users. Then, we extract the technologies (e.g., tags) from SO posts. Questions of SO have tags (e.g., Fig. \ref{fig:technology-example}). Thus, we can directly extract the discussed technologies in the questions utilizing the specified tags. However, the answers of SO do not specify any tags. Intuitively, the technologies of answers are the same as the questions for which they are submitted. Therefore, we extract the answer technologies from their questions. In the SO dataset, each answer has an attribute named \emph{ParentId} (i.e., question ID of that answer). We use ParentId to extract the target question and the technologies for answers. 

We then attempt to determine which technologies a user has worked on for over 15 years. First, we group each user's posts according to the submission year. Next, we extract the set of technologies from each year's posts and distribute them over the years. We then collect all the tags that represent any versions or components of our selected 160 technologies from the SO data dump. We find $9,141$ tags that are related to these 160 technologies. For example, we find 581 tags that are connected to JavaScript (all the tags related to our 160 technologies can be found in our online appendix \cite{replicationPackage}). We also consider the tag synonyms. Please note that the Stack Exchange data dump offers tag synonyms~\cite{datadumpapi}. For example, ``c-sharp" is a synonym of c\#. We keep tags extracted from posts if they match any of our 160 technologies or synonyms or their related ones. Later, we replaced the synonyms or related tags with the corresponding standard technologies (i.e., any of our 160 technologies). 
We kept the entries blank when Q\&A footprints of a particular user were not found in any of the years.

To compute the average usage span, we first list all the technologies $T_i$, where $i = {1, 2, 3, \ldots, K}$. Here, $K$ is the total number of technology. We then compute the average usage span (in years) of each of the listed technologies $U_{avg}$ using the following formula.
    
    \begin{equation}
        U_{avg} = \frac{1}{n} \sum_{u = 1}^{N} \sum_{y = 1}^{Y} T_i
    \end{equation}
    
\noindent where $Y$ represents the total number of years (i.e., $15$), $N$ represents the total number of selected users (i.e., $96,786$), and $n$ represents the total number of users who use the currently selected technology for at least one year. The value of $T_i$ is $1$ when the technology belongs to the set of technologies in a particular year and $0$ otherwise. Consider the technology distribution shown in Table \ref{table:technology-usage-per-user}. The value of $T_i$ is $1$ for 2008, 2009, 2011, 2013 \& 2015 and $0$ for the remaining years when the selected technology is c\#. Thus, the usage span of c\# for the selected user can be calculated as 
    $U_{avg} (c\#) = \frac{1}{1} \sum_{u = 1}^{1} \sum_{y = 1}^{10} c\# = \frac{1}{1} (5) = 5\hspace{1mm}\text{years}$.
We calculate the average usage span of each of the technologies and finally report the top 50 technologies with a higher usage span than others.
    
\textbf{Finding Emerging Technologies.} In this step, we attempt to find the emerging technologies to which more users have switched over the years. We first calculate the total number of users working on a particular technology each year. Then we examine whether the number of users has increased over the years. For example, let the number of users for two consecutive years of a particular technology be $U_{Y_{i}}$ and $U_{Y_{i+1}}$, respectively. 
Then $U_{Y_{i+1}} - U_{Y_{i}} \geqslant 0$ implies that the number of users has increased over the year. Such an increase indicates that the users prefer these technologies more over time since the number of users we selected is fixed. We list the technologies as emerging ones when the above condition is true for all the consecutive years we selected.

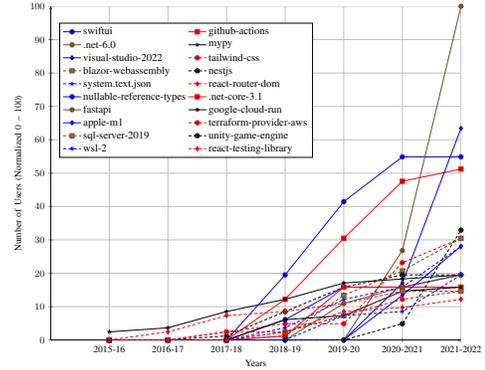
\begin{figure}[!htb]
\centering
  \resizebox{2.5in}{!}{%
    \begin{tikzpicture}
    \begin{axis}[
        width=6in,
        height=5in,
        axis lines=left,
        grid=both,
        legend cell align=left,
        legend style={
            anchor=north,
            legend columns=2,
            at={(0.33,0.95)}
        },
        unbounded coords=jump,
        label style={font=\small},
        ticklabel style={font=\small},
        ymin=0,
        xmin=0,
        ymax=100,
        xmax=7,
        xtick={1,2,3,4,5,6,7},
        xticklabels={2015-16, 2016-17, 2017-18, 2018-19, 2019-20, 2020-2021, 2021-2022},
        ytick={0,10,...,100},
        xlabel=Years,
        ylabel=Number of Users (Normalized $0$ -- $100$),
        legend entries={
        swiftui,
        github-actions,
        .net-6.0,
        mypy,
        visual-studio-2022,
        tailwind-css,
        blazor-webassembly,
        nestjs,
        system.text.json,
        react-router-dom,
        nullable-reference-types,
        .net-core-3.1,
        fastapi,
        google-cloud-run,
        apple-m1,
        terraform-provider-aws,
        sql-server-2019,
        unity-game-engine,
        wsl-2,
        react-testing-library
        },
        cycle list name = auto 
    ]
    \addplot coordinates {(1,0)	(2,0)	(3,0)	(4,19.51219512)	(5,41.46341463)	(6,54.87804878)	(7,54.87804878)};
    \addplot coordinates {(1,0)	(2,0)	(3,0)	(4,12.19512195)	(5,30.48780488)	(6,47.56097561)	(7,51.2195122)};
    \addplot coordinates {(1,0)	(2,0)	(3,0)	(4,0)	(5,0)	(6,26.82926829)	(7,100)};
    \addplot coordinates {(1,2.43902439)	(2,3.658536585)	(3,8.536585366)	(4,12.19512195)	(5,17.07317073)	(6,18.29268293)	(7,19.51219512)};
    \addplot coordinates {(1,0)	(2,0)	(3,0)	(4,0)	(5,0)	(6,17.07317073)	(7,63.41463415)};
    \addplot coordinates {(1,0)	(2,0)	(3,2.43902439)	(4,4.87804878)	(5,4.87804878)	(6,23.17073171)	(7,30.48780488)};
    \addplot coordinates {(1,0)	(2,0)	(3,0)	(4,1.219512195)	(5,13.41463415)	(6,20.73170732)	(7,30.48780488)};
    \addplot coordinates {(1,0)	(2,0)	(3,1.219512195)	(4,8.536585366)	(5,15.85365854)	(6,19.51219512)	(7,19.51219512)};
    \addplot coordinates {(1,0)	(2,0)	(3,0)	(4,3.658536585)	(5,12.19512195)	(6,15.85365854)	(7,28.04878049)};
    \addplot coordinates {(1,0)	(2,2.43902439)	(3,7.317073171)	(4,8.536585366)	(5,10.97560976)	(6,14.63414634)	(7,15.85365854)};
    \addplot coordinates {(1,0)	(2,0)	(3,0)	(4,6.097560976)	(5,15.85365854)	(6,15.85365854)	(7,19.51219512)};
    \addplot coordinates {(1,0)	(2,0)	(3,0)	(4,1.219512195)	(5,15.85365854)	(6,15.85365854)	(7,15.85365854)};
    \addplot coordinates {(1,0)	(2,0)	(3,0)	(4,0)	(5,10.97560976)	(6,15.85365854)	(7,19.51219512)};
    \addplot coordinates {(1,0)	(2,0)	(3,0)	(4,6.097560976)	(5,7.317073171)	(6,14.63414634)	(7,15.85365854)};
    \addplot coordinates {(1,0)	(2,0)	(3,0)	(4,0)	(5,0)	(6,13.41463415)	(7,28.04878049)};
    \addplot coordinates {(1,0)	(2,0)	(3,2.43902439)	(4,2.43902439)	(5,7.317073171)	(6,12.19512195)	(7,14.63414634)};
    \addplot coordinates {(1,0)	(2,0)	(3,0)	(4,2.43902439)	(5,7.317073171)	(6,14.63414634)	(7,14.63414634)};
    \addplot coordinates {(1,0)	(2,0)	(3,0)	(4,0)	(5,0)	(6,4.87804878)	(7,32.92682927)};
    \addplot coordinates {(1,0)	(2,0)	(3,0)	(4,0)	(5,7.317073171)	(6,8.536585366)	(7,19.51219512)};
    \addplot coordinates {(1,0)	(2,0)	(3,0)	(4,2.43902439)	(5,8.536585366)	(6,9.756097561)	(7,12.19512195)};
    
    \end{axis}
    \end{tikzpicture}
    }
\caption{Emerging technologies.}
\vspace{-3mm}
\label{fig:emerging-technology}
\end{figure}

\section{Study Findings} \label{sec:studyResults}

\subsection{Technology with High Usage span (RQ1)} \label{subsec:techUsability}

Fig. \ref{fig:technology-usability} shows the top 50 technologies with a high usage span.
Seventeen of them are related to programming/scripting/style sheet/markup language. According to our investigated Q\&A traces, JavaScript has the highest usage span. On average, developers have used JavaScript for more than four years. Three more languages, such as C\#, Java, and Python, have more than three years of usage span. Among the selected 50, 10 technologies are associated with software frameworks. They have 1.6--2.9 years of usage span. For example, ASP.NET frameworks have the highest 2.9 years of usage span. We get five commonly used Operating Systems in the top 50 technologies list. However, iOS, Windows, and Android have more usage span (2.2, 2.2 \& 2.1 years) over Linux (1.9 years) and macOS (1.7 years) operating systems.

Our selected list contains one development platform with a 1.8-year usage span. Besides, we get the most used version control named git in our top 50 high-usage technology list. 
Furthermore, the integrated development environment visual studio was also found to have 2.4 years of usage span.
Among the remaining technologies, five were database \& relational database management systems (e.g., MySQL) with 1.7--2.2 years of usage span, two query languages (e.g., SQL) with 2.3 \& 1.7 years of usage span, two compilers with 1.6 years of usage span and five other technologies (e.g., Google Service) with 1.2--2.0 years of usage span on average. 

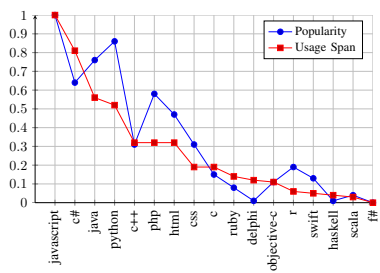
\begin{figure}[pt]
\centering
  \resizebox{2in}{!}{%
    \begin{tikzpicture}
    \begin{axis}[
        width=4in,
        height=2.5in,
        axis lines=left,
        grid=both,
        legend cell align=left,
        legend style={
        	font=\small,
        	legend pos=north east,
        	legend cell align=left
        },
        xmin=0,
        ymax=1,
        xmax=17,
        xtick={1,2,3,4,5,6,7,8,9,10,11,12,13,14,15,16,17},
        xticklabels={javascript,
                    c\#,
                    java,
                    python,
                    c++,
                    php,
                    html,
                    css,
                    c,
                    ruby,
                    delphi,
                    objective-c,
                    r,
                    swift,
                    haskell,
                    scala,
                    f\#
                    },
        x tick label style={rotate=90,anchor=east},
        ytick={0,0.1,0.2,...,1},
        legend entries={Popularity, 
        Usage Span,
        },
        cycle list name = auto 
    ]

    \addplot
    coordinates {(1,1)
                (2,0.64)
                (3,0.76)
                (4,0.86)
                (5,0.31)
                (6,0.58)
                (7,0.47)
                (8,0.31)
                (9,0.15)
                (10,0.08)
                (11,0.01)
                (12,0.11)
                (13,0.19)
                (14,0.13)
                (15,0.01)
                (16,0.04)
                (17,0)
                };
    
    \addplot
    coordinates {(1,1)
                (2,0.81)
                (3,0.56)
                (4,0.52)
                (5,0.32)
                (6,0.32)
                (7,0.32)
                (8,0.19)
                (9,0.19)
                (10,0.14)
                (11,0.12)
                (12,0.11)
                (13,0.06)
                (14,0.05)
                (15,0.04)
                (16,0.03)
                (17,0)
                };

    \end{axis}
    \end{tikzpicture}
    }
\caption{Popularity vs usage span.
}
\vspace{-3mm}
\label{fig:popularity-vs-usability}
\end{figure}


\subsection{Emerging Technology (RQ2)} \label{subsec:emergingTech}

Fig. \ref{fig:emerging-technology} shows the top 20 emerging technologies to which more of our selected users switched to work over the years. We find that more users have started using frameworks like SwiftUI. It offers a reliable user interface toolkit for iOS, iPadOS, watchOS, tvOS, and macOS and helps developers design apps in a declarative way. Besides, the .NET-6.0 platform and Visual Studio 2022 IDE  also emerged over time. .NET-6.0 is a unified development platform allowing developers to build applications for the cloud, web, desktop, mobile, gaming, IoT, and AI. .NET 6 and Visual Studio 2022 provide hot reload, new git tooling, intelligent code editing, robust diagnostics and testing tools, and better team collaboration. Blazor WebAssembly framework built by Microsoft that allows developers to build single-page web applications. We find that FastAPI, a modern, high-performance web framework for building APIs with Python 3.7+, has grown sharply over the years. In addition, the NestJS framework, Google's Cloud Run, and SQL Server 2019 (which comes with Apache Spark and Hadoop Distributed File System) have attracted more users over the years. 

\section{Discussion}
\label{sec:discussion}

We attempt to see whether the popularity agrees with the usage span of technologies. In particular, we compare the popularity and usage span of the 17 programming/scripting/style sheet/markup languages (Fig. \ref{fig:technology-usability}). In SO, the popularity of technology is typically estimated by how many questions tagged that technology. We thus count the questions for each of the 17 languages \cite{datadumpapi}. Finally, we normalize the number of questions and average usage years from 0 to 1 for each technology. 
Fig. \ref{fig:popularity-vs-usability} shows the line graph popularity vs. usage span. We see that popularity does not always guarantee a high usage span. For example, Python is more popular than C\#. However, C\# has a higher usage span on average than Python. Similarly, PHP has more popularity than C++. Interestingly, they have the same usage span. 
Such findings suggest that choosing a technology based on only its popularity does not always guarantee a high usage span.

\section{Implications} 
\label{sec:implications}

Our investigation and findings could guide novice developers, startup software companies, and software clients to select technologies with high usability usage. Computer science (or information technology) educators and researchers could also benefit from our study.

$\bullet$ \textbf{Novice Developers.} 
The software technology field progresses rapidly with numerous alternatives for a single task \cite{chen2016mining}. Therefore, it is crucial to choose appropriate technologies, especially for novice developers. They often look for popular ones. However, we find that only popularity is not always a good measure since it does not guarantee a high usage span. Alternatively, they can seek suggestions from senior developers. However, such suggestions from any individual could be biased by the developer's personal preference. Thus, mining technologies with high usage span by analyzing 
SO 
could reliably guide novice developers in selecting technologies. 

$\bullet$ \textbf{Startup Software Companies.} 
Investors always prefer profitable business. Thus, startup software companies conduct costly market analyses to identify the appropriate technologies they should start. One of their main objectives is to lower maintenance costs because about 90\% of software life cost is associated with its maintenance phase \cite{dehaghani2013factors, granja1997method}. However, technologies with a high usage span could be a better choice. It can be assumed that the technologies developers use for a long time, their maintenance is easy and cost-effective.

$\bullet$ \textbf{Software Clients.} 
Software clients can be individuals or organizations who usually lack in-depth technical knowledge. However, many software comes with a maintenance agreement for a specific time (e.g., a few months). Upon expiration, owners have to maintain the purchased software. Thus, clients must also be aware of the appropriate technology to keep maintenance costs low while buying software. Study findings on technologies with a high usage span could help clients choose programming language and database management systems that developers prefer to use for a long time. 
choice 
of technologies may reduce overall maintenance (e.g., bug fixing, resolving security issues).

$\bullet$ \textbf{Educators in Computer Science (or Information Technology).} 
Educators (e.g., faculty members and instructors) often guide classroom students in selecting technologies. They also prepare/revise the course curriculum. Introducing stable technologies in the curriculum and teaching them in the classroom might make the students fit for the industry. 

$\bullet$ \textbf{Software Engineering Researchers.} 
We initiate 
an investigation that attempts to find technologies with a high usage span. 
The results may inspire researchers to conduct further explorations in the future and introduce tool supports to assist developers in estimating usage and other quality metrics.

\section{Limitations \& Future Research Scopes} 
\label{sec:limitations}

We estimate the usage span and find emerging technologies based on the Q\&A footprints of SO. However, we analyzed the footprints who created their profile in 2008 and 2009. Including more users' Q\&A footprints could help solidify the findings. Furthermore, including data from other crowd-sourced developers' forums (e.g., GitHub) might increase generalizability and confidence in our findings.

We distributed the technologies for each user over the years (e.g., Table \ref{table:technology-usage-per-user}). However, in some cases, we did not find any questions or answers from users for a whole year. Therefore, we kept those entries blank, assuming that the developers were not active in their development during those years. However, it is also possible they were active but did not post questions or submit answers. Such a dilemma might cause computing errors in calculating the average usage years of technologies. 
When we simply remove the samples with one or multiple blank entries and calculate the usage span, we find 42 technologies common in the top 50. However, more filtration techniques could be applied to compute usage span, which can enhance confidence in the results. 

\section{Related Work} 
\label{sec:relatedWork}

Several existing studies investigate the technology landscape \cite{chen2016mining, chen2016techland}, trends \cite{kim2008visualization, park2019gitviz, barua2014developers, woon2009framework} and skills \cite{montandon2020skills} utilizing crowd-sourced knowledge. To the best of our knowledge, we first investigate the usage span of technologies.

Chen and Xing~\cite{chen2016mining} mine the technology landscape from SO question tags and present such landscape as a graphical technology associative network. Chen et al.~\cite{chen2016techland} present TechLand, a system for assisting technology landscape inquiries with categorical and relational knowledge of technologies. However, TechLand does not offer any information about the usage span of those technologies.
Kim et al.~\cite{kim2008visualization} collect keywords from patent documents, cluster the documents, and then form a semantic network of keywords from those documents. A patent map is then built by reordering each keyword node of the semantic network. This study enables the viewers to understand emerging technologies. Yoon and Magee~\cite{yoon2018exploring} utilize patent information (e.g., citations) to discover promising technology. Our study also offers a set of emerging technologies. However, the context is different, and we focus on software technologies.

Park et al.~\cite{park2019gitviz} explore and visualize technology trends. First, they use data from the GitHub repository to identify the key technologies and developers in a particular field. Then they identify other technologies related to that field and explore the changes in popularity of those technologies over time. Barua et al.~\cite{barua2014developers} discover topics from the developers' discussions of SO. Then they analyze the relationships and popularity trends of those technologies over time. However, we argue that popularity does not always guarantee a high usage span. 
Baquero et al.~\cite{baquero2017predicting} investigate relationships between programming languages utilizing SO information. In particular, they extract information related to 18 programming languages and find the languages close to each other (e.g., Matlab and R). 

Montandon et al.~\cite{montandon2020skills} investigate hard and soft skills demanded in IT companies. They analyze the description of 20K job opportunities posted on the SO job site. They find that programming languages are the most demanded hard skills. We investigate the usage span of technologies and report the top 50 technologies with a high usage span, where 17 are programming languages. 

Zou et al.~\cite{zou2015non} analyze the non-functional requirements from SO discussions. They find that developers focus most on usability and reliability. Our study investigates the usage span of technologies utilizing developers' Q\&A traces of SO over 15 years that was not investigated in previous studies.

\section{Conclusion}
\label{conclusion}
Technology selection is crucial for developers among a number of similar options for a particular task. Choosing inappropriate technologies might impede developers' career growth. We thus analyze about 97K developers' 6.7 million Q\&A footprints over 15 years to determine the usage span of technologies. We report the top 50 technologies with a high usage span that contains -- (1) 17 programming languages, (2) 10 software frameworks, (3) four operating systems, (4) one development platform, (5) one version control system, (6) two IDEs, (7) five database 
\& database management systems, (8) two query languages, (9) two compilers, and (10) five others. We also attempt to find the emerging technologies and report 20 of them. Such emerging technologies to which more developers are switching over time include SwiftUI, .NET-6.0 \& Blazor WebAssembly frameworks, and Visual Studio 2022 IDE.
To validate our findings, we plan to -- (1) include more developers' footprints from SO and other crowd-sourced developer forums and (2) survey developers to find their agreement on our findings in the future.

\bibliographystyle{plainnat}
{\footnotesize
\bibliography{reference}}

\end{document}